\documentclass{emulateapj}
 \usepackage{url}

\shorttitle{Statistical Distributions of Optical Flares}
\shortauthors{Yi et al}
\slugcomment{ }

\begin{document}

\title{Statistical Distributions of Optical Flares from Gamma-Ray Bursts}
\author{Shuang-Xi Yi$^{1,3, 4}$, Hai Yu$^{2,4}$, F. Y. Wang$^{2,4}$, and Zi-Gao Dai$^{2,4}$}
\affil{$^{1}$College of Physics and Engineering, Qufu Normal University, Qufu 273165, China; \\
       $^{2}$School of Astronomy and Space Science, Nanjing University, Nanjing 210093, China; fayinwang@nju.edu.cn\\
       $^{3}$GXU-NAOC Center for Astrophysics and Space Sciences, Department of Physics, Guangxi University, Nanning 530004;\\
       $^{4}$Key laboratory of Modern Astronomy and Astrophysics (Nanjing University), Nanjing 210093, China.\\
       }

\begin{abstract}
We statistically study gamma-ray burst (GRB) optical flares from the
{\em Swift}/UVOT catalog. We compile 119 optical flares,
including 77 flares with redshift measurements. Some tight
correlations among the time scales of optical flares are found. For
example, the rise time is correlated with the decay time, and the
duration time is correlated with the peak time of optical flares.
These two tight correlations indicate that longer rise times are
associated with longer decay times of optical flares, and also
suggest that broader optical flares peak at later times, which are
consistent with the corresponding correlations of X-ray flares. We
also study the frequency distributions of optical flare parameters,
including the duration time, rise time, decay time, peak time and
waiting time. Similar power-law distributions for optical and X-ray
flares are found. Our statistic results imply that GRB optical
flares and X-ray flares may share the similar physical origin and
both of them are possibly related to central engine activities.
\end{abstract}

\section{Introduction}\label{sec:intro}

Gamma-ray bursts (GRBs) are the most luminous phenomena occurring at
cosmological distances. It is well known that the prompt gamma-ray
emission is produced by internal dissipation processes within the
relativistic ejecta (Piran 2004; M{\'e}sz{\'a}ros 2006; Zhang 2007;
Kumar \& Zhang 2015), while the broadband afterglows are usually
interpreted as the interaction of an ultra-relativistic ejecta with
the ambient medium (M{\'e}sz{\'a}ros \& Rees 1997; Sari 1998). The
successful launch of the {\em Swift} satellite in 2004 (Gehrels et
al. 2004) has greatly improved our understanding of GRB physics.
Since its rapid response, {\em Swift} could quickly allow the X-Ray
Telescope (XRT) and the Ultraviolet/Optical Telescope (UVOT) to
localize a GRB position and begin to observe the afterglow (Burrows
et al. 2005a; Roming et al. 2005).

Some new phenomena are also discovered in the {\em Swift} satellite
era, while the most intriguing phenomenon is the erratic flares of
the ``canonical'' X-ray light curve, observed in the early X-ray
afterglow phase (Burrows et al. 2005b; Zhang et al. 2006; Nousek et
al. 2006). Those erratic X-ray flares usually happen at $\sim
10^2-10^5$\,s after the prompt emission (Falcone et al. 2007;
Chincarini et al. 2007, 2010; Swenson \& Roming 2014), and are
observed in both long and short GRBs (Romano et al. 2006; Falcone et
al. 2006; Campana et al. 2006; Margutti et al. 2011). Since flares
appear to come from a distinct emission mechanism than the
underlying afterglow emission, and are seen in both long and short
GRBs, it is generally supposed to be powered by the central engine
activities. Therefore, X-ray flares and the prompt gamma-ray
emission may have the similar physical origins (Burrows et al.
2005b; Fan \& Wei 2005a; Falcone et al. 2006, 2007; Zhang et al.
2006; Nousek et al. 2006; Wu et al. 2006; Chincarini et al. 2007,
2010; Abdo et al. 2011; Troja et al. 2015; Yi et al. 2015; Mu et al.
2016a). Interestingly, flares also appear in the UV/optical band. Li
et al. (2012) selected a group of optical light curves with flares,
and suggested that optical flares are also related to the erratic
behavior of the central engine, which are similar to X-ray flares.
Flares are both observed in the X-ray as well as the UV/optical
bands, but the number of GRBs with optical flares is much smaller
than that of GRBs with X-ray flares.

Flares are common astrophysical phenomena throughout the universe.
Some studies on X-ray flares from astrophysical systems have been
carried out (Wang et al. 2015). Wang \& Dai (2013) selected 83 GRB
X-ray flares and 11595 solar X-ray flares, and performed a
statistical comparison between them. They found the energy,
duration, and waiting-time distributions of GRB X-ray flares are
similar to those of solar flares, which suggest a similar physical
origin of the two kinds of flares. Some works using different
methods and data also obtain a similar result (Aschwanden 2011; Wang
et al. 2015; Harko et al. 2015; Guidorzi et al. 2015). These results
are supported by Yi et al. (2016), who studied all significant X-ray
flares from GRBs observed by {\em Swift} until March 2015, and
obtained 468 bright X-ray flares, including 200 flares with
redshifts. They found that there are four power-law distributions
with different indices between X-ray flares and solar flares,
including power-law distributions of energies, durations, peak
fluxes and waiting times. Besides, they also studied the peak times,
rising times and decay times of GRB X-ray flares, and found all of
them show the power-law distributions. These similar statistical
distributions between solar flares and GRB X-ray flares suggest both
of them could be produced by magnetic reconnection, and also could
be explained by a fractal diffusive, self-organized criticality
model (Aschwanden 2011; Wang \& Dai 2013; Harko et al. 2015; D{\u
a}nil{\u a} et al. 2015; Yi et al. 2016). Interestingly, some
theoretical models have been proposed that GRB X-ray flares could be
powered by magnetic reconnection events (Giannios 2006; Dai et al.
2006; Zhang \& Yan 2011; Mu et al. 2016b).

In this paper, we investigate the optical flares observed by {\em
Swift/UVOT} and study the distributions of optical flare parameters,
such as duration times, rise times, decay times, peak times and
waiting times. Since optical flares and X-ray flares may have a
common physical origin, both of them may have similar distributions
of the parameters. This paper is organized as follows. In Section 2,
we present the selected GRB sample. In section 3, we study some
correlations between parameters of optical flares. The distributions
of flare parameters are discussed in Section 4. Discussion is given
in Section 5. Section 6 presents conclusions. A concordance
cosmology with parameters $H_0 = 71$ km s $^{-1}$ Mpc$^{-1}$,
$\Omega_M=0.30$, and $\Omega_{\Lambda}=0.70$ is adopted in all part
of this work.

\section{Data}
We extensively search for the optical flares of GRBs. Since the
fraction of GRBs with optical flares is much smaller than that of
X-ray flares, we mainly focus on GRBs detected by {\em Swift/UVOT}.
Swenson et al. (2013) carefully studied the second UVOT GRB
afterglow catalog, which provides a complete data set of fitted UVOT
light curves for both long and short GRBs observed by {\em Swift}
from 2005 April through 2010 December (Roming et al. 2009). They
found more than one hundred unique potential flares in 68 different
optical light curves, and obtained the starting time, peak
time and end time of optical flares. We consider the full sample
containing 119 optical flares (see their Table 2), including 77
flares with redshifts. These optical flares usually contain a
complete structure, including remarkable rising and decaying phase.
Figure 1 shows an sample of the optical flares. Most GRBs have a
single optical flare, but some of them have several optical flares.

We carefully study the timescales of optical flares, such as
waiting time, duration time, peak time, rise time and decay time.
The time parameters of optical flares are derived as follows, which
also can be seen in Yi et al. (2016). The rise time can be derived
by $T_{rise}=T_{peak}-T_{start}$, the decay time
$T_{decay}=T_{stop}-T_{peak}$ and the duration time
$T_{Duration}=T_{stop}-T_{start}$, where $T_{start}$, $T_{peak}$ and
$T_{stop}$ are the starting time, peak time and end time of flares,
respectively. They are all listed in Table 2 of Swenson et al.
(2013). The waiting time for one flare is defined as
$T_{waiting}=T_{start,i+1}-T_{start,i}$, where $T_{start,i+1}$ is
the observed start time of the $i + 1th$ flare, and $T_{start,i}$ is
the observed start time of the $ith$ flare. All the optical flare
properties should be transferred into the source rest frame, if they
have redshift measurements in the following analysis. For the first
flare appearing in an optical afterglow, the rest-frame waiting time is
simply taken as $T_{start}/(1 + z)$, where $z$ is redshift. We next
study the frequency distributions of the duration time, waiting time,
rise time, decay time and peak time of optical flares. Since optical
flares and X-ray flares may have a common physical origin, we will
compare the results of optical flares with X-ray flares, and check
whether both of them show similar distributions of parameters.

\section{Parameters of optical Flares and Correlations}
Figure 2 shows the waiting time and peak time histogram
distributions of the optical flares. The waiting times of optical
flares range from 10 s to $10^6$ s after the GRB trigger, mainly from
$10^2$ s to $10^3$ s, which is similar to the distribution of X-ray
flares. The peak time $T_{peak}$ of optical flare is in the range of
$10^2$ s to $10^6$ s, mainly in $10^2$ s to $10^3$ s, occurring at
the early time of the optical afterglow, which is consistent with
the peak time distribution of X-ray flares. The optical smooth onset
bumps are also peaking at the early time of the afterglow, but the
optical flares and onset bumps are different from each other.
According to the standard forward shock model, the onset of GRB
afterglow is characterized by a smooth bump in the early afterglow
when the ultra-relativistic fireball is decelerated by the
circumburst medium, and these features are well consistent with the
forward shock models (Molinari et al. 2007; Liang et al. 2010, 2013; Yi et
al. 2013).

We use the simple linear regression analysis for parameter
fitting\footnote{https://en.wikipedia.org/wiki/Simple\_linear\_regression}
(Chatterjee et al. 2000), which is a linear regression model with a
single variable. In this paper, we only consider correlations
between two parameters of optical flares. So the simple linear
regression method is adequate. Interestingly, these correlations
have been found in X-ray flares of GRBs (Chincarini et al. 2007,
2010; Yi et al. 2016). Therefore, similar correlations could be
expected in the optical flares. If treating multivariate correlations,
multiple regression method should be used. Figure 3
presents the strong correlation between the rise and decay times of
the optical flares. The rise time is tightly correlated with the
decay time with the slope index of 0.99. There is also a strong
correlation between the duration time and the peak time, with the
slope index of 1.11. These two strong correlations indicate that
longer rise times are associated with longer decay times for optical
flares, and also suggest that broader optical flares peak at later
times. The two correlations of optical flares are in good agreement
with the corresponding correlations of X-ray flares, which can be
seen from Figure 3 of Yi et al. (2016). The best fitting results of
the four correlations are shown in Table 1. These tight correlations
suggest that the structures of the optical and X-ray flares are
similar, indicating a similar physical origin of them. Besides,
Figure 3 also exhibits the correlations between the waiting time and
other parameters of the optical flares. The waiting time is
correlated with both the peak time and the duration time of optical
flares, which indicates a longer waiting time tends to peak at a
later time with a longer duration time, which are also consistent
with the correlations of X-ray flares. In the fitting, the
errors of parameters are not considered.
Because there are no parameter errors reported in the optical flares from the
second UVOT GRB Catalog (Swenson et al. 2013). In order to test whether the fitting results are biased by parameter errors,
we assume the errors are randomly changed in the $10 - 20 \%$ range of the original
value. We take the $T_{waiting}-T_{peak}$ correlation as an example. After considering the parameter errors, we find that the best-fitted power-law index is
0.90 using the method proposed by Kelly (2007). So this result is consistent with that in Table 1.

However, the start and end times of flares are affected by
the observational temporal gaps in the Swift light curve, so the
time scales of flares, such as, the rise time, decay time and
duration time, may be also changed. In order to test this
observational bias for the four correlations, we simulate $10^4$
times for each correlation. In each simulations, the time scales
will be randomly changed in the $0 - 10 \%$ range of the original
value. Then we refit the correlations. The best-fitting results are
shown in Figure 4. From this figure, we can see that the fitting
results from simulations are slightly different from the value
derived from observation data. They are consistent with each other
at $1\sigma$ confidence level.

Some other instrumental and observational biases, which tend
to disfavor flares with short durations at late times or smooth
flares with long durations, will affect the observational
correlations. Because the UVOT collects data in event mode during
the first $\sim$1000 s, while later observations are performed in
image mode. The former mode has full temporal resolution, the latter
mode integrates light over the whole exposure. For this reason,
flares with a duration of 100-200 s cannot be easily detected at
late times, because the observations average the emission over
several hundreds of seconds. Furthermore, the identification of a
flare requires also the identification of the underlying continuum.
A smooth, longer-lasting flare could be more easily misclassified as
continuum and therefore missed. Overall, the temporal correlations
of optical flares may derive from instrumental and observational
biases, especially for those flares with short durations but at late
time, or smooth flares with long durations. We take the
$T_{Duration}-T_{peak}$ correlation as an example. In order to test
the two biases, we provide two groups of simulation data in Figure
5, i.e., optical flares with short durations at late times for the
first bias (the red circles), and the smooth flares with long
duration times for the second bias (the blue circles). For the first
bias, we simulate 500 optical flares with $100 s <T_{Duration}< 300
s$ and $10^5 s<T_{peak}<10^6 s$. For the second bias, we simulate
500 optical flares with $5\times10^4 s <T_{peak}< 10^6 s$ and
$5\times10^4 s<T_{Duration}<10^6 s$. The simulated red points of the
first bias are far away from the best fitting line in Figure 5,
implying the correlation may be affected by instrumental bias. For
the second bias, if smooth, longer-lasting flares could be
identified, because peak time and duration time are almost on the
same order of magnitude, they marginally follows the
$T_{Duration}-T_{peak}$ correlation.

\section{The Frequency Distributions of Flare Parameters}

In this section, we use the maximum likelihood estimation
method (Bauke 2007) to fit the frequency distributions of optical
flare parameters, such as $T_{waiting}$, $T_{Duration}$,
$T_{peak}$, $T_{rise}$ and $T_{decay}$. We investigate the
differential distributions of parameters for the optical flares. The
differential distribution is chosen as a power-law form
\begin{equation}\label{differentialFun}
    \rho(x)=\beta x^{\alpha_x},
\end{equation}
where $\alpha_x$ is the power-law index. The occurrence rate of
flares in each bin can be calculated from $\rho(x)=N/\Delta x$,
where $N$ and $\Delta x$ are the number of flares in the bin, and
the width of the bin, respectively.

We have studied the frequency distributions of X-ray flares in Yi et
al. (2016). We focus on the optical flares observed by {\em Swift}
in this paper. There are 119 GRB optical flares in our sample, and
77 of them have redshifts which consist of a sub-sample. For the
total sample, we divide the parameter $x$, which represents the
parameter of flares, into 11 bins. Figure 6 shows the differential
distributions of the optical flare in the total sample. The points
are the observed data with $1\sigma$ errors, which are chosen as the
Poisson error. The red curves are the optimal fittings derived with
the maximum likelihood estimation method. The optimal fitted
parameters $\alpha_x$ for the distributions of the waiting time,
duration time, peak time, rise time and decay time are
$1.24\pm0.08$, $1.23\pm0.07$, $1.28\pm0.09$, $1.31\pm0.10$ and
$1.21\pm0.07$, respectively. The sub-sample is treated with the
similar method after the redshift correction, and the optimal
fitting results are shown in Figure 7. The optimal fitted parameters
$\alpha_x$ for the distributions of waiting time, duration time,
peak time, rise time and decay time are $1.30\pm0.11$,
$1.29\pm0.09$, $1.29\pm0.10$, $1.27\pm0.10$ and $1.28\pm0.11$,
respectively. Figures 6 and 7 show that the differential
distributions of the flare parameters can be well described by
power-law functions. We find that both optical flares and X-ray
flares have similar statistical distributions, so we suppose optical
flares and X-ray flares have a common physical origin, which implies
that both of them may be powered by activities of central engines.

\section{Discussion}
X-ray flares are the most common phenomena in GRB X-ray
afterglows. According to the statistical results of X-ray flares,
about more than one-third of {\em Swift} GRBs with remarkable
flares. However, the number of flares in the UV/optical are far less
than those of X-ray. Therefore, not all the optical flares
correspond to X-ray flares. Swenson et al. (2013) applied the
Bayesian Information Criterion to analyze the residuals of the
fitted UV/optical light curve, and identified 119 unique flaring
periods. In this paper, we study the properties of optical flares,
comparing them with X-ray flares. We check all the optical light
curves of Swenson et al. (2013) and X-ray afterglows. We find that
most of GRBs in this sample have notable flares simultaneously
observed in the X-ray band, but there are still about a dozen GRBs
with no distinct flare activities in X-ray band.

The temporal behaviors of flares are different from the
underlying afterglow emissions, however they are well consistent
with those of prompt gamma-ray emissions. Therefore, X-ray and
optical flares are supposed to be produced by internal emission
powered by central engine. Through a comparative of the afterglow
observations, there is evidence suggesting that the optical and
X-ray flares originate from similar physical processes (Swenson et
al. 2010; Li et al. 2012). Most optical flares usually happen at
early time after the prompt emission. However, some flares are even
occurring at very late time both in the X-ray and the UV/optical,
such as GRB 070318 and GRB 090926A. Flares of these two bursts are
not only observed at early time, but matching well at late time
greater than $10^5$ s in both bands. Therefore, the physical origin
of both optical and X-ray flares may be similar, and both of them
are related to central engine. However, as discussed above, the
presence of flares in one band, but not in another is usually seen.
We suppose the primary reason is the lower significance of most
flares in the lower energy bands. While the X-ray flares are often
easily identified by visual inspection of the light curves,
potential optical flares are more often overlooked or dismissed as
noise (Swenson et al. 2010; Li et al. 2012; Swenson \& Roming 2013).
Whether X-ray and optical flares have the same origin remains an
important open question, much more observation data are required to
answer this question.

Interestingly, other bumps are also occurring in the optical afterglows.
But they are different from optical flares. Generally, the onset of GRB
afterglow is seen by a smooth bump in early optical afterglow light curve
as the fireball shell is decelerated by the circumburst medium. Liang
et al. (2010) extensively searched for the afterglow onset bump
feature from early afterglow light curves, and 20 optical onset bumps are identified.
These optical afterglows have smooth bumps, with the rising index for most GRBs
is $1-2$, and the decay index is $0.44-1.77$. These afterglow onset features are
well consistent with the external-forward shock (FS) model. Another sharp optical bump
is produced by the reverse shock (RS) emission. However, the RS emission is rarely
appeared in the optical afterglows. At present, only a small fraction of GRBs shows RS emission in
optical afterglows. According to the RS model, the theoretical rising index can be steep as 5
in the thin shell case for a constant interstellar medium (Kobayashi 2000, Yi et al. 2013). One interesting case is GRB 041219A,
which shows three significantly power-law rise and fall peaks in the optical-IR band. The
first optical peak tracking the gamma-ray light curve during the prompt emission can be
understood as emission from internal shocks, while the remaining two peaks are produced
by RS and FS component, respectively (Blake et al. 2005; Fan et al. 2005b). Another similar burst is GRB 110205A (Gendre et al. 2012).
Therefore, the two optical bumps (FS component and RS component)
are attributed to the external shock emission, while the optical flares are related to
the internal shocks of the central engine. In our analysis, such bumps are not included.

\section{Conclusions}
In this paper, we have complied 119 optical flares of GRBs taken
from {\em Swift/UVOT} catalog of Swenson et al. (2013) until
December 2010, including 77 flares with redshifts. We studied the
parameters of the optical flares, such as the waiting time, duration
time, rise time and decay time. We found the waiting times of
optical flares range from between 10 s and $10^6$ s after the GRB
trigger, and the peak time of an optical flare is in the range of
$10^2$ s to $10^6$ s. We also found some tight correlations between
these time scales of optical flares. Generally, these tight
correlations suggest that longer rise times associate with longer
decay times, and also suggest broader optical flares peak at later
times. These properties are consistent with the results of X-ray
flares, and indicate the structures of optical flares and X-ray
flares are similar. However, these correlations may be
affected by the instrumental bias, e.g., flares with short durations
but at late time are hard to identify by UVOT. We also studied the
frequency distributions of the parameters of optical flares. The
best-fitting results for the power-law distributions of the
parameters of the optical flares are similar with those of X-ray
flares. Our results indicate GRB optical flares and X-ray flares may
share the similar physical origin, and both of them are related to
central engine activities.


\acknowledgments

We thank the anonymous referee for useful comments and suggestions.
We thank En-Wei Liang, Xue-Feng Wu and Jie-Shuang Wang for useful comments and helps.
This work is supported by the National Basic Research Program of
China (973 Program, grant No. 2014CB845800), the National Natural
Science Foundation of China (grants 11422325, 11373022, and
11573014), the Excellent Youth Foundation of Jiangsu Province
(BK20140016), China Postdoctoral
science foundation (grant No. 2017M612233), Science and Technology Program of QuFu Normal
University (xkj201614).

\begin{figure*}
\centering
\includegraphics[angle=0,scale=0.32]{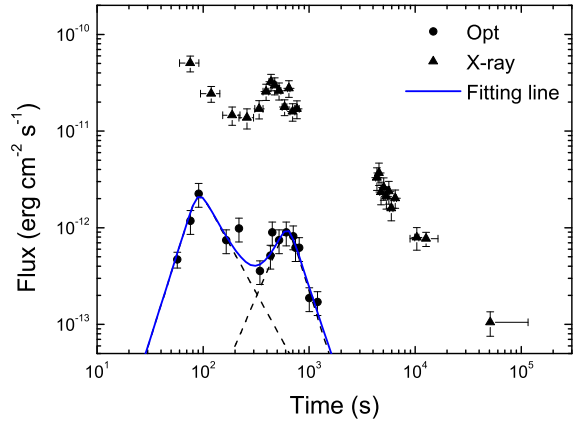}
\caption{The sample of two optical flares of GRB 060926. The two optical cases show the remarkable flare features, which is also corresponding to an X-ray flare, simultaneously.}
\end{figure*}

\begin{figure*}
\includegraphics[angle=0,scale=0.30]{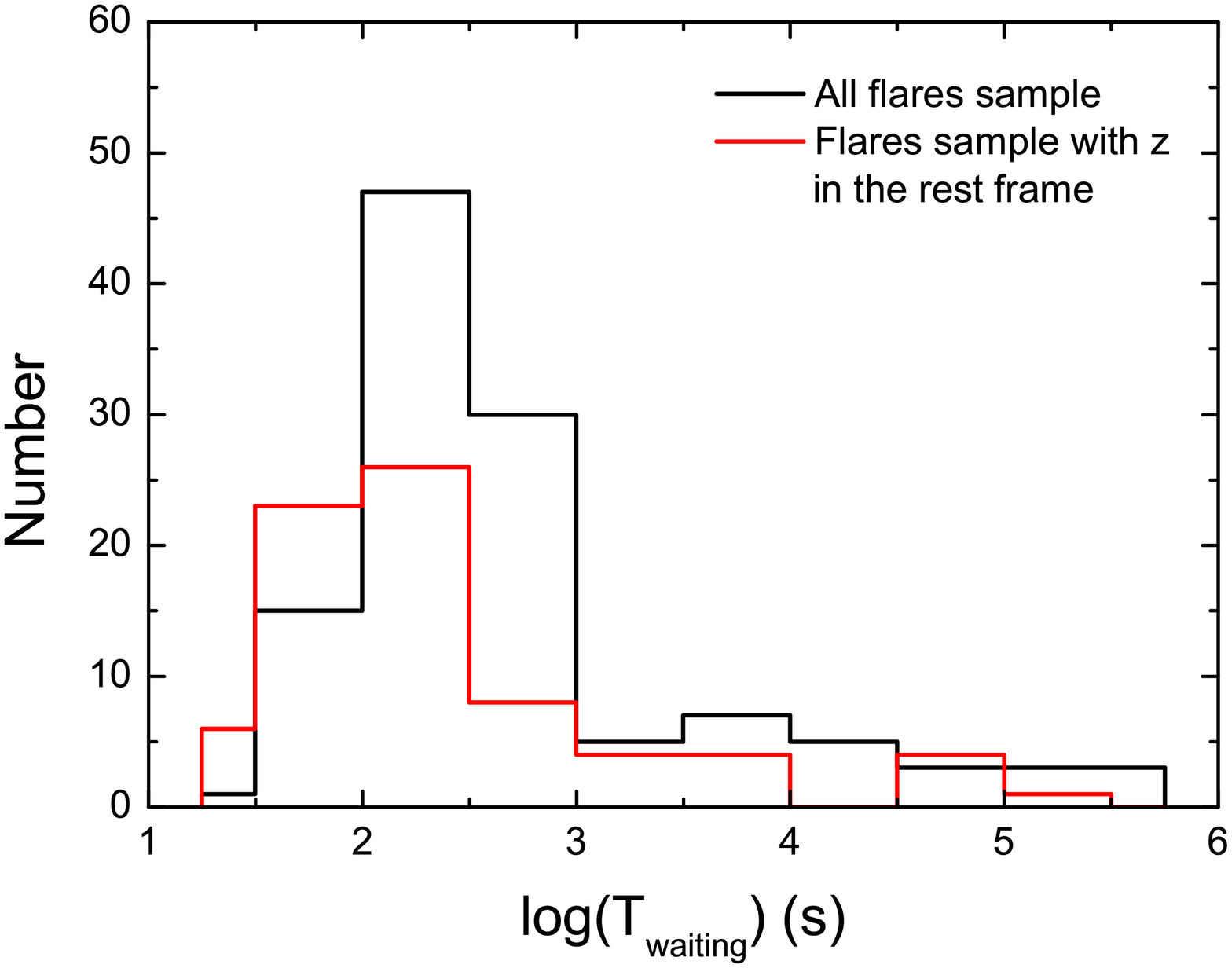}
\includegraphics[angle=0,scale=0.30]{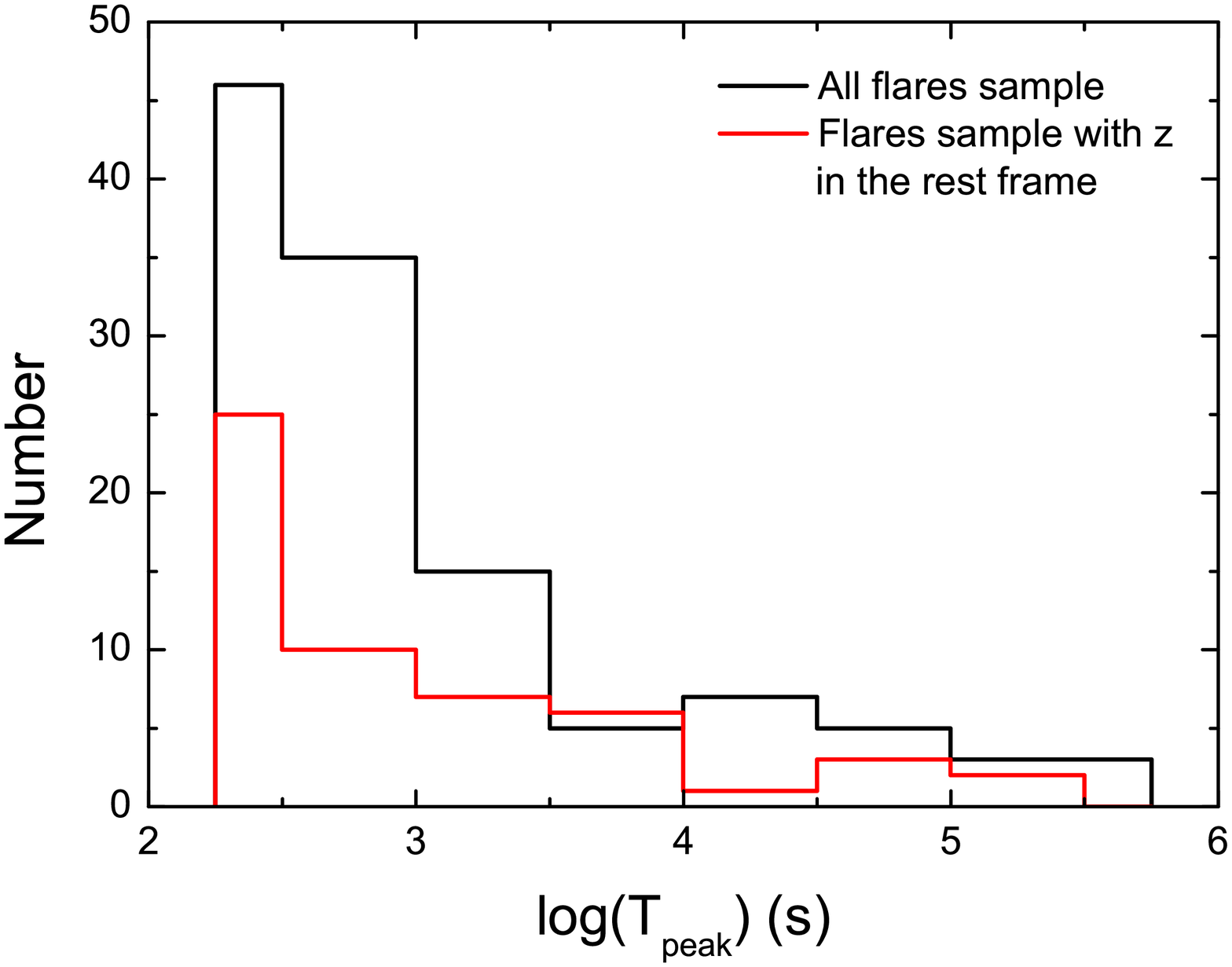}
\caption{The waiting time and peak time distributions of optical
flares. The black line is corresponding to all of the sample. The
red line represents the flares with reshifts, and the parameters
have been transferred to the source frame. }
\end{figure*}

\begin{figure*}
\includegraphics[angle=0,scale=0.30]{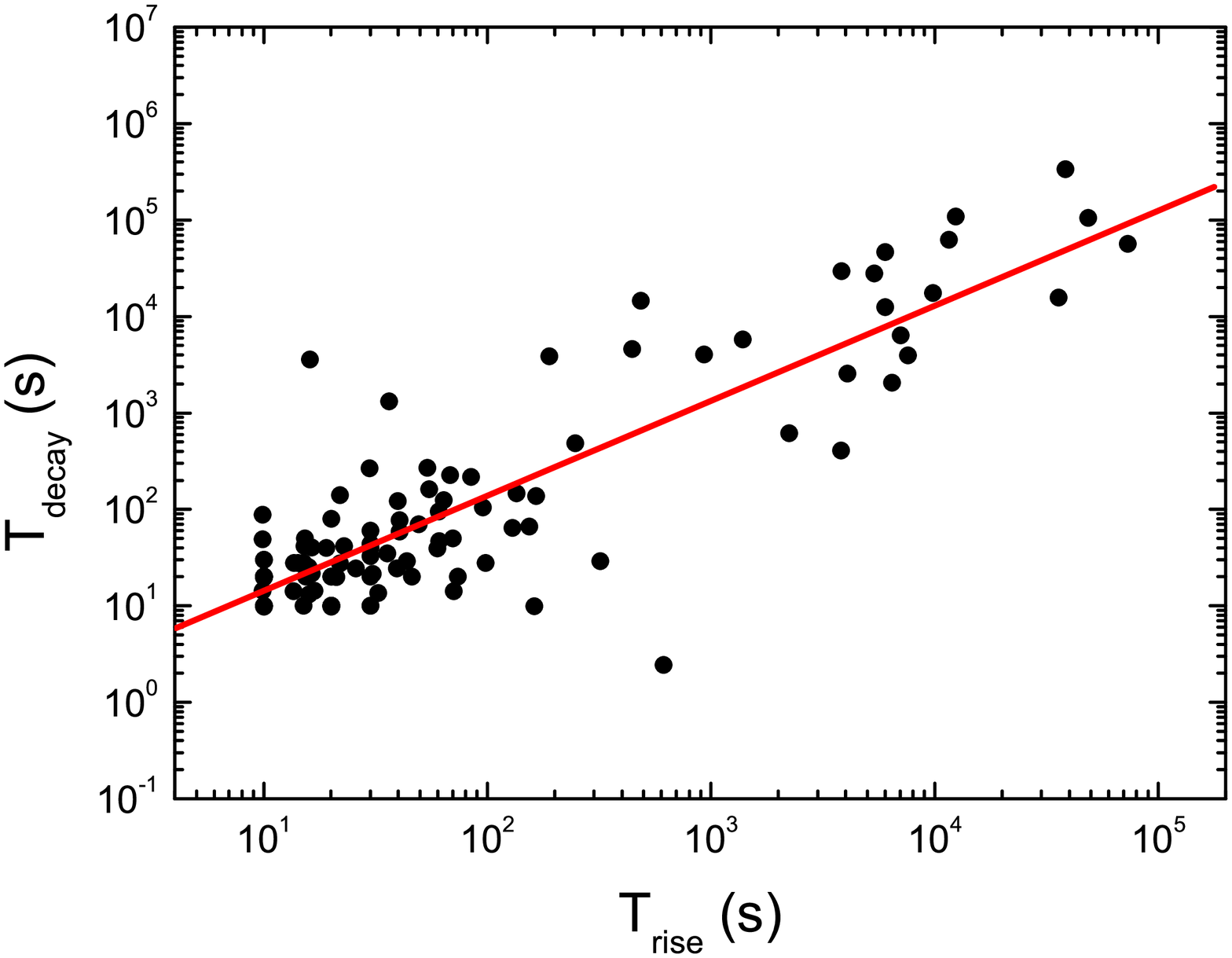}
\includegraphics[angle=0,scale=0.30]{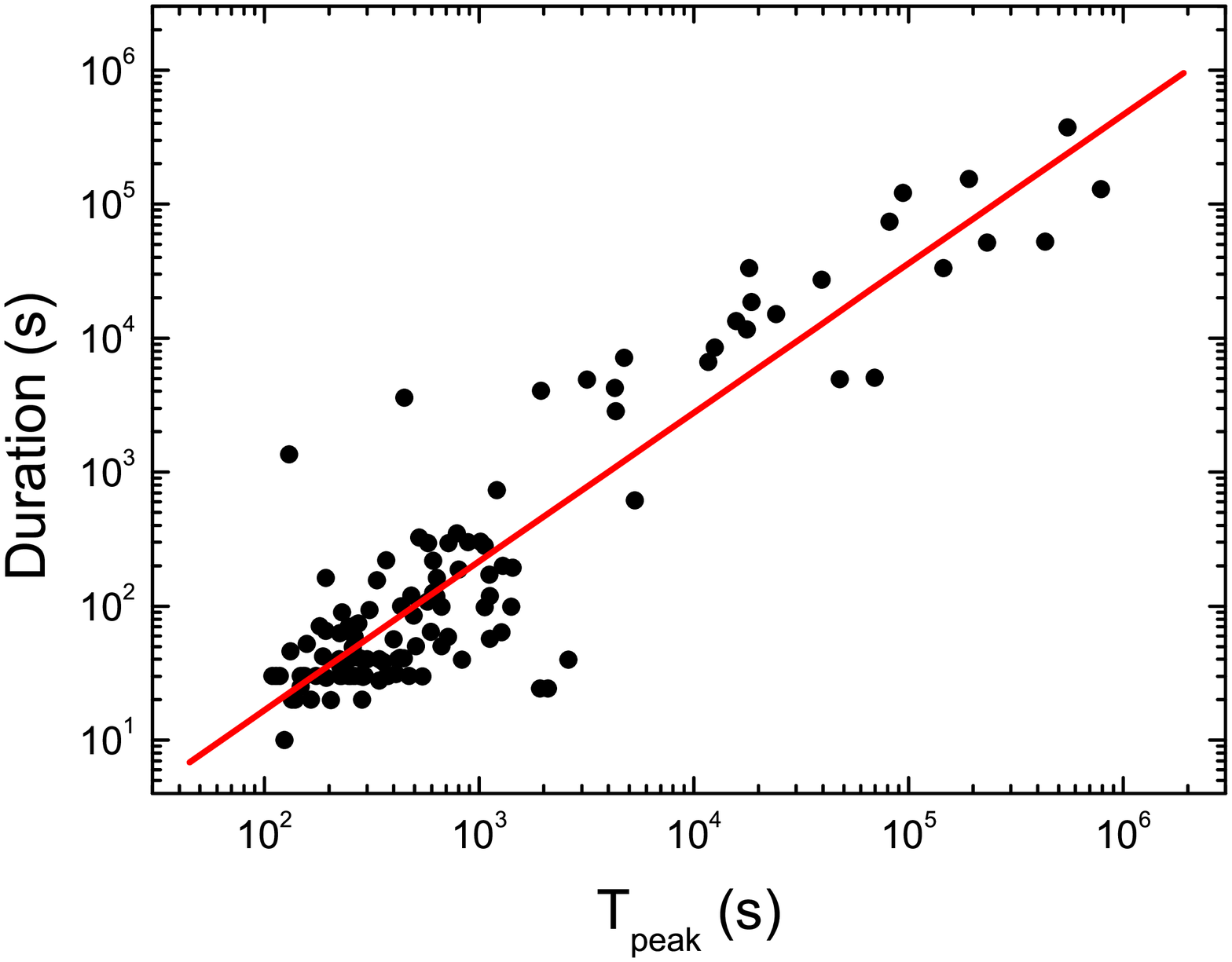}
\includegraphics[angle=0,scale=0.30]{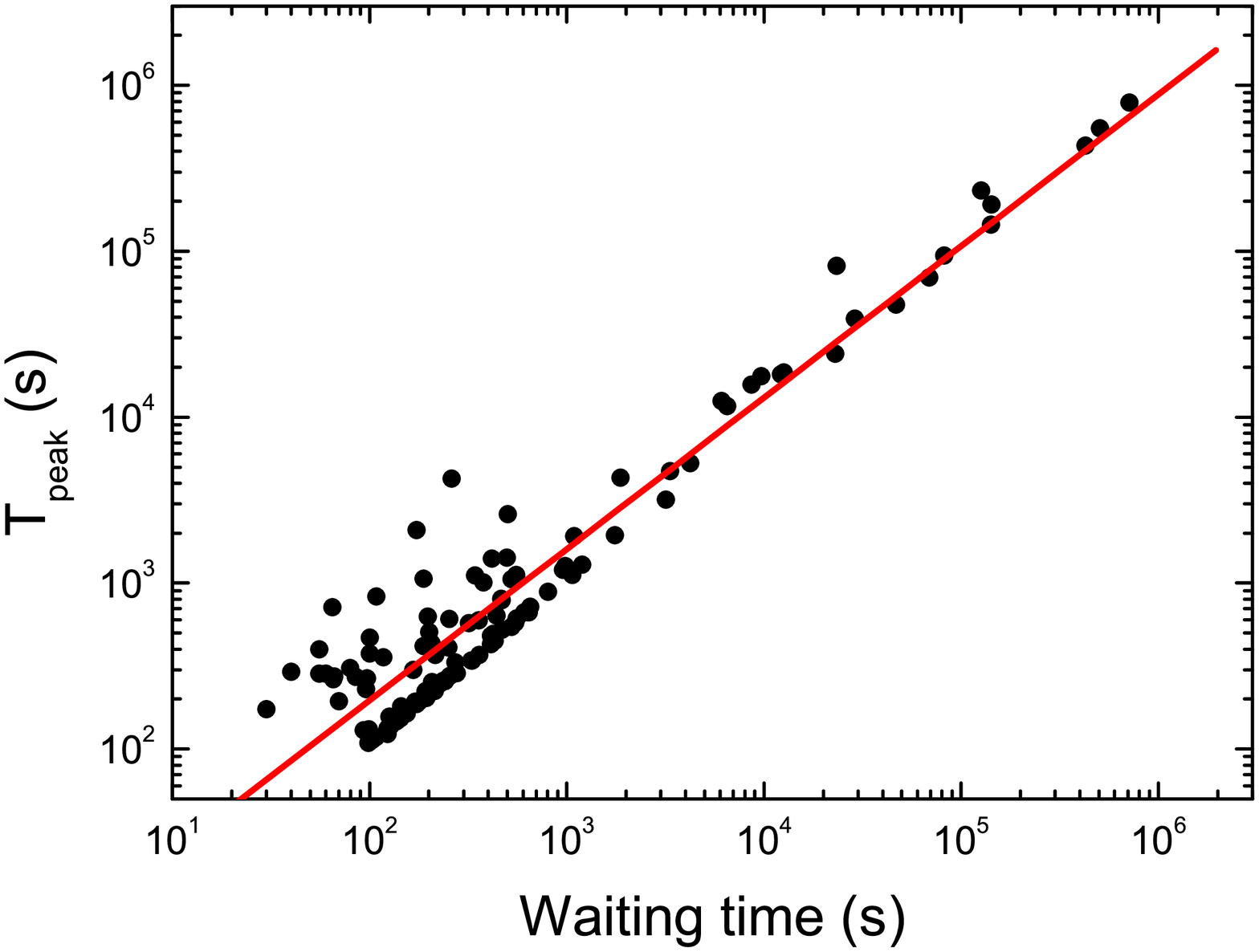}
\includegraphics[angle=0,scale=0.30]{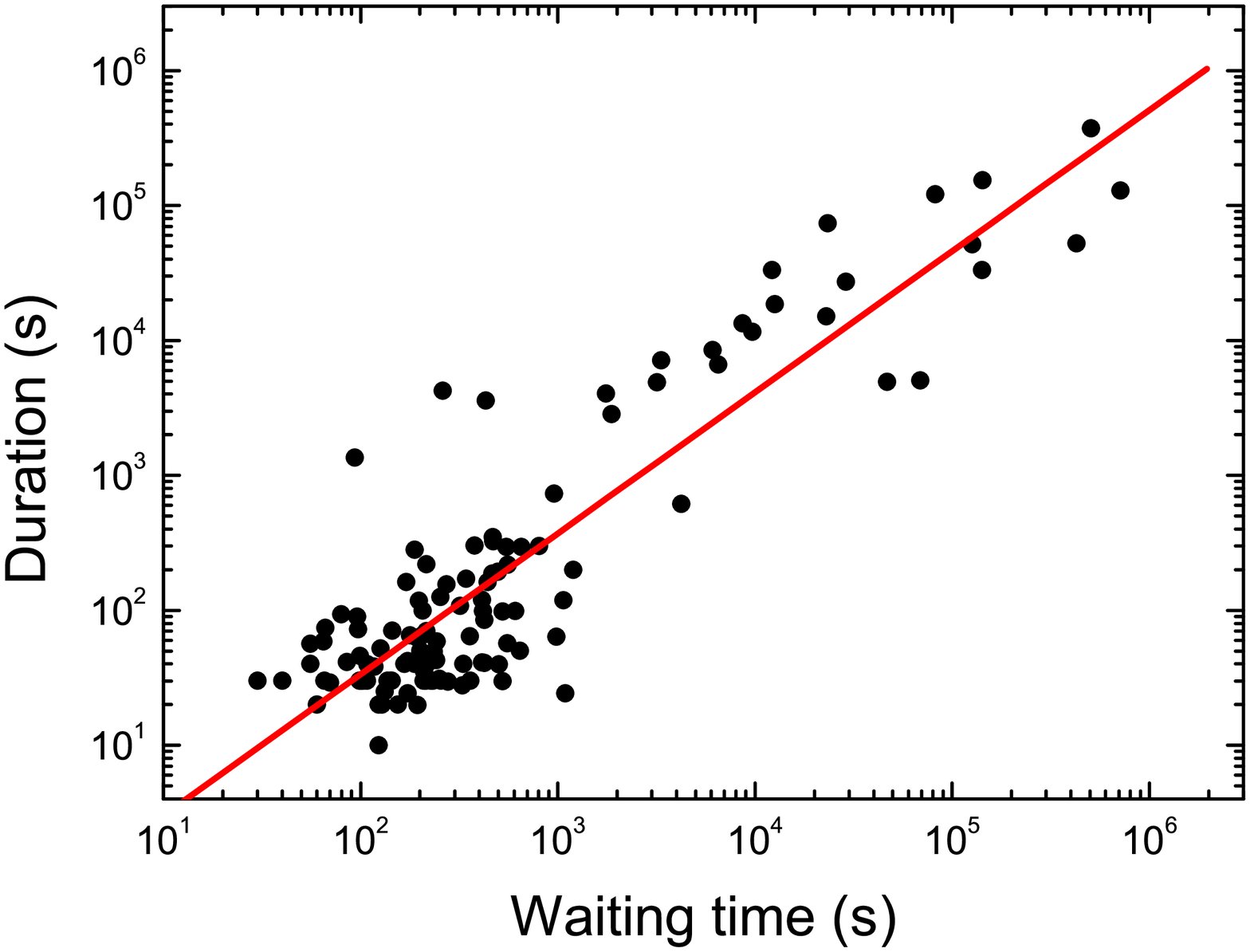}
\caption{The correlations between time scales of GRB optical flares.
The rise time is correlated with the decay time, and the duration is
correlated with the peak time, which are consistent with the results of
X-ray flares. The red line is the best fitting, which is shown in Table 1.}
\end{figure*}

\begin{figure*}
\includegraphics[angle=0,scale=0.30]{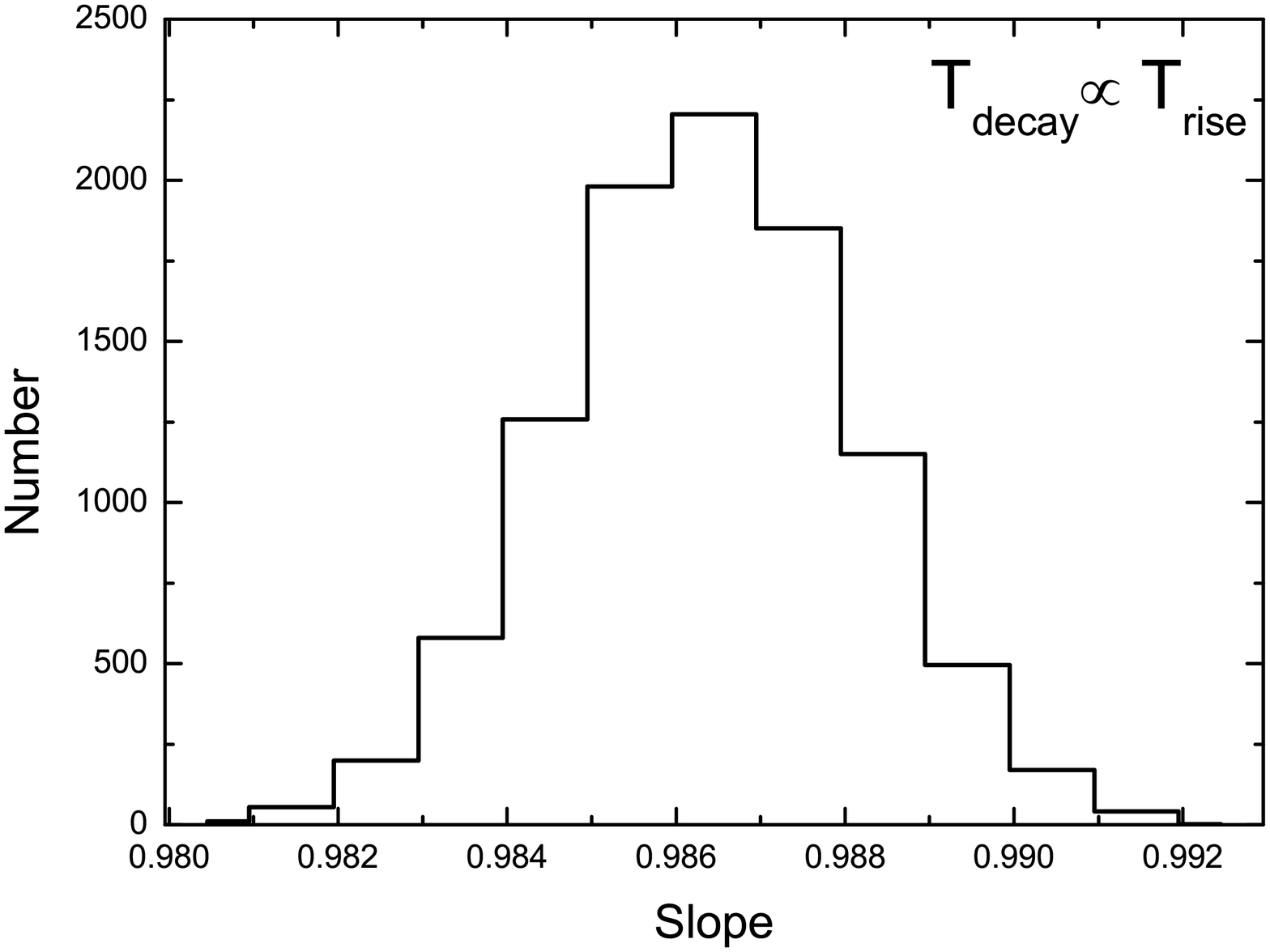}
\includegraphics[angle=0,scale=0.37]{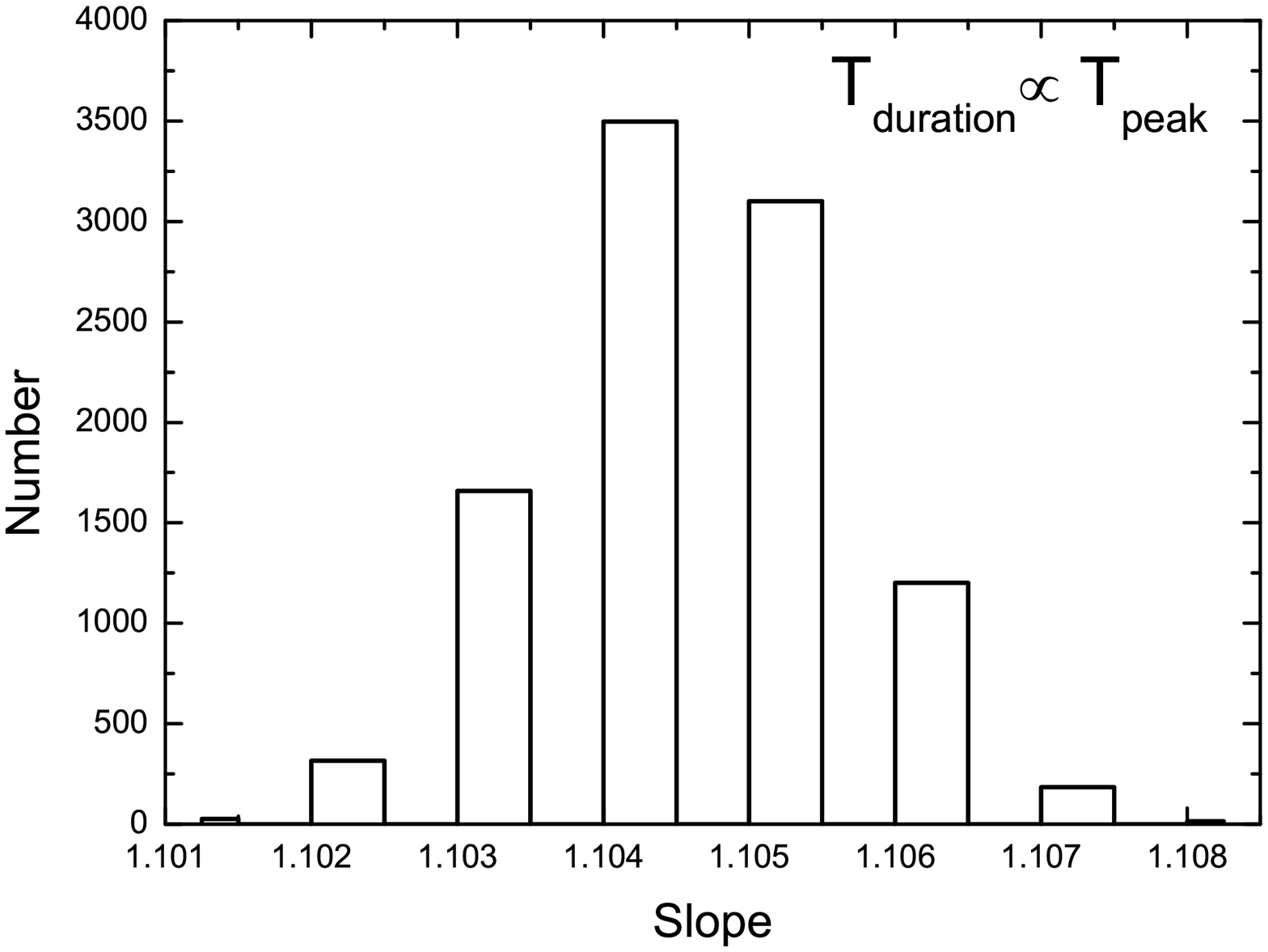}
\includegraphics[angle=0,scale=0.30]{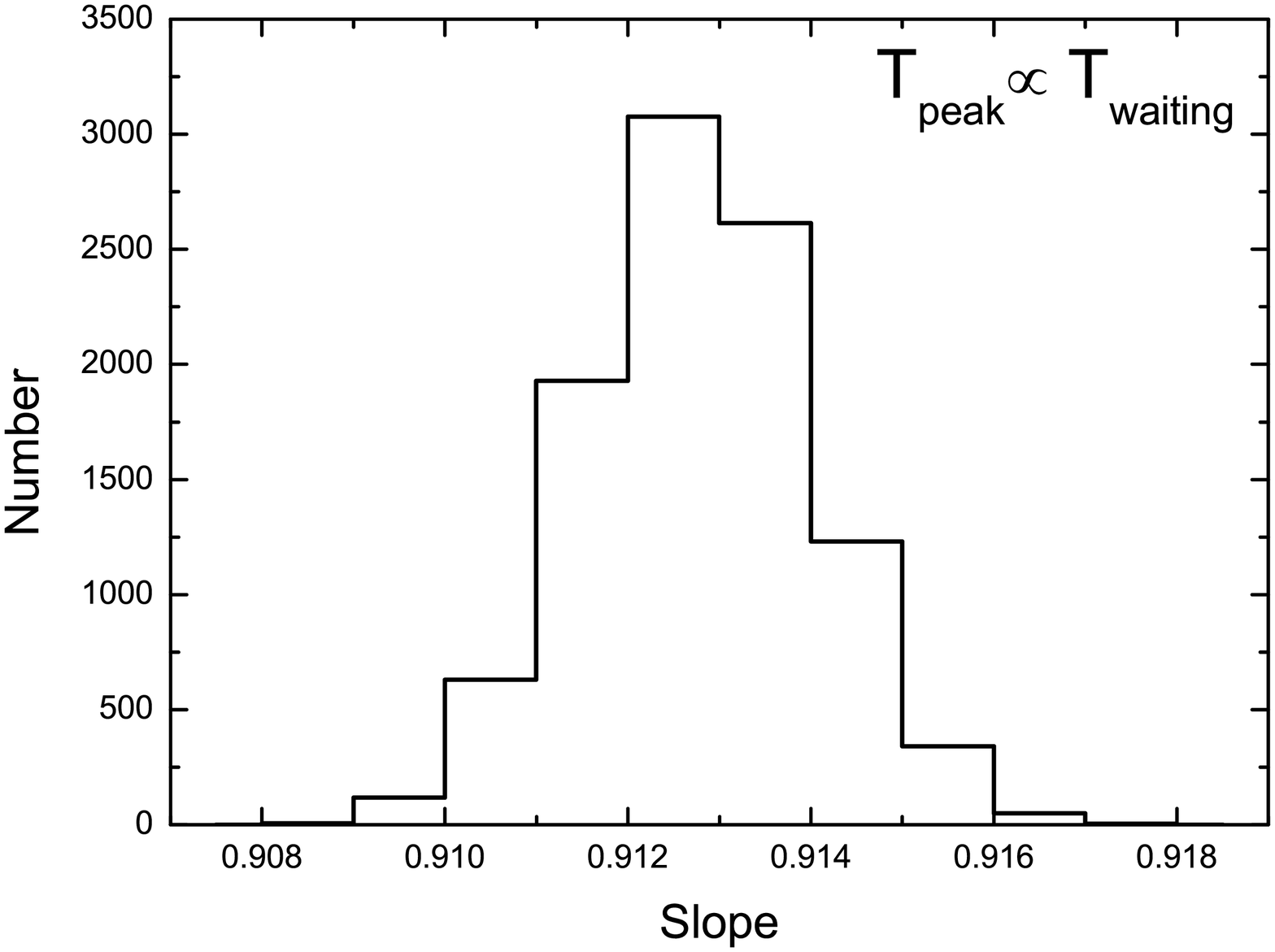}
\includegraphics[angle=0,scale=0.30]{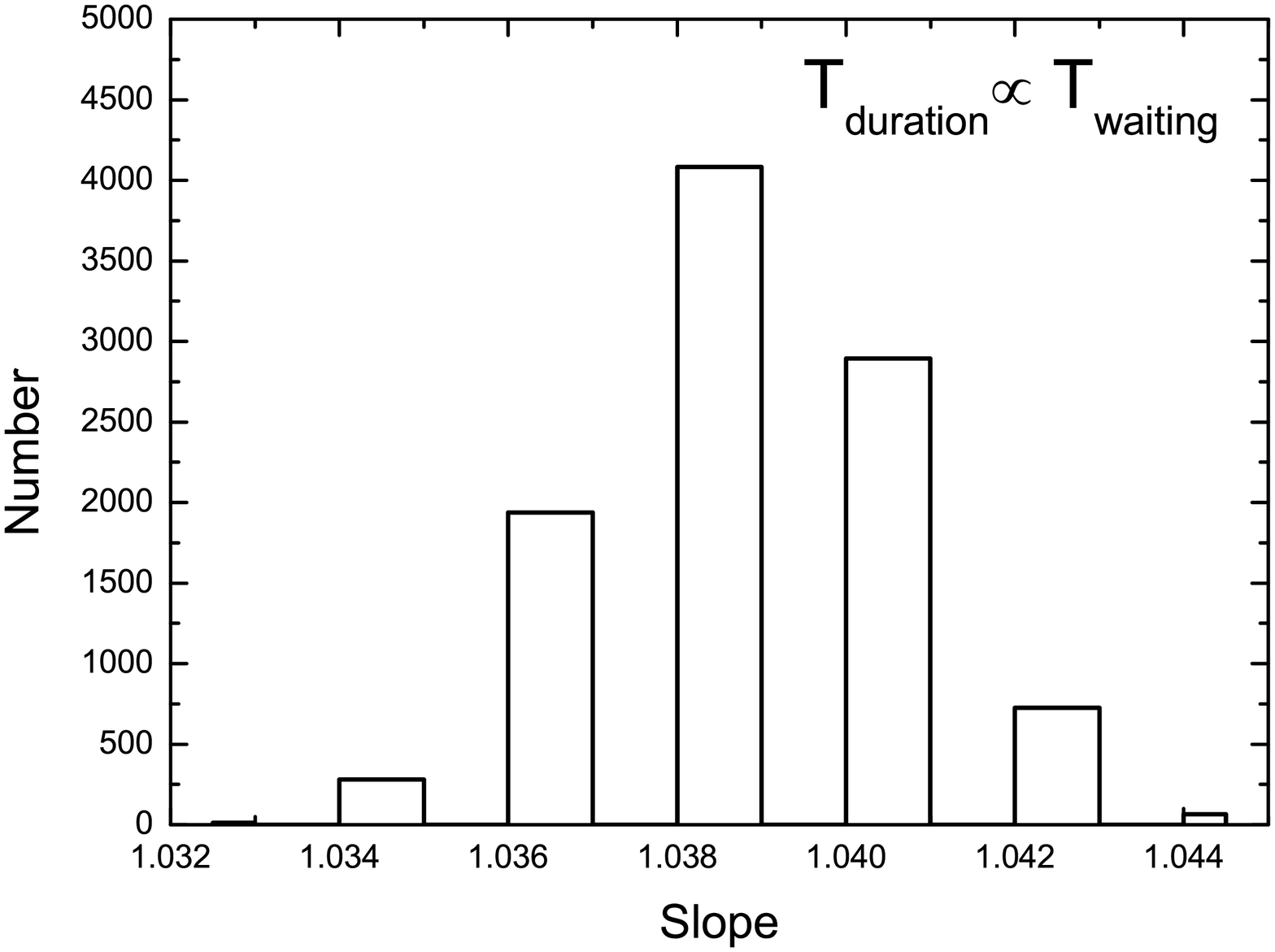}
\caption{The distributions of the best fitting results for the four
correlations from simulation data. We simulate $10^4$ times for each
correlation.}
\end{figure*}

\begin{figure*}
\centering
\includegraphics[angle=0,scale=0.30]{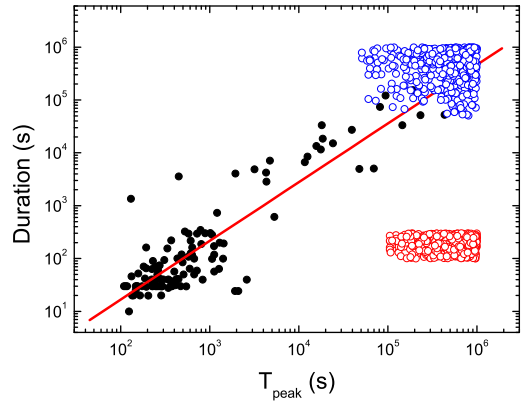}
\caption{Two simulations for the instrumental and observational
biases for the $T_{Duration}-T_{peak}$ correlation: flares with
short duration times at late times (the red circles) and smooth,
longer-lasting flares (the blue circles). It's obviously that the
instrumental bias is significant. }
\end{figure*}

\begin{figure*}
\includegraphics[angle=0,scale=0.30]{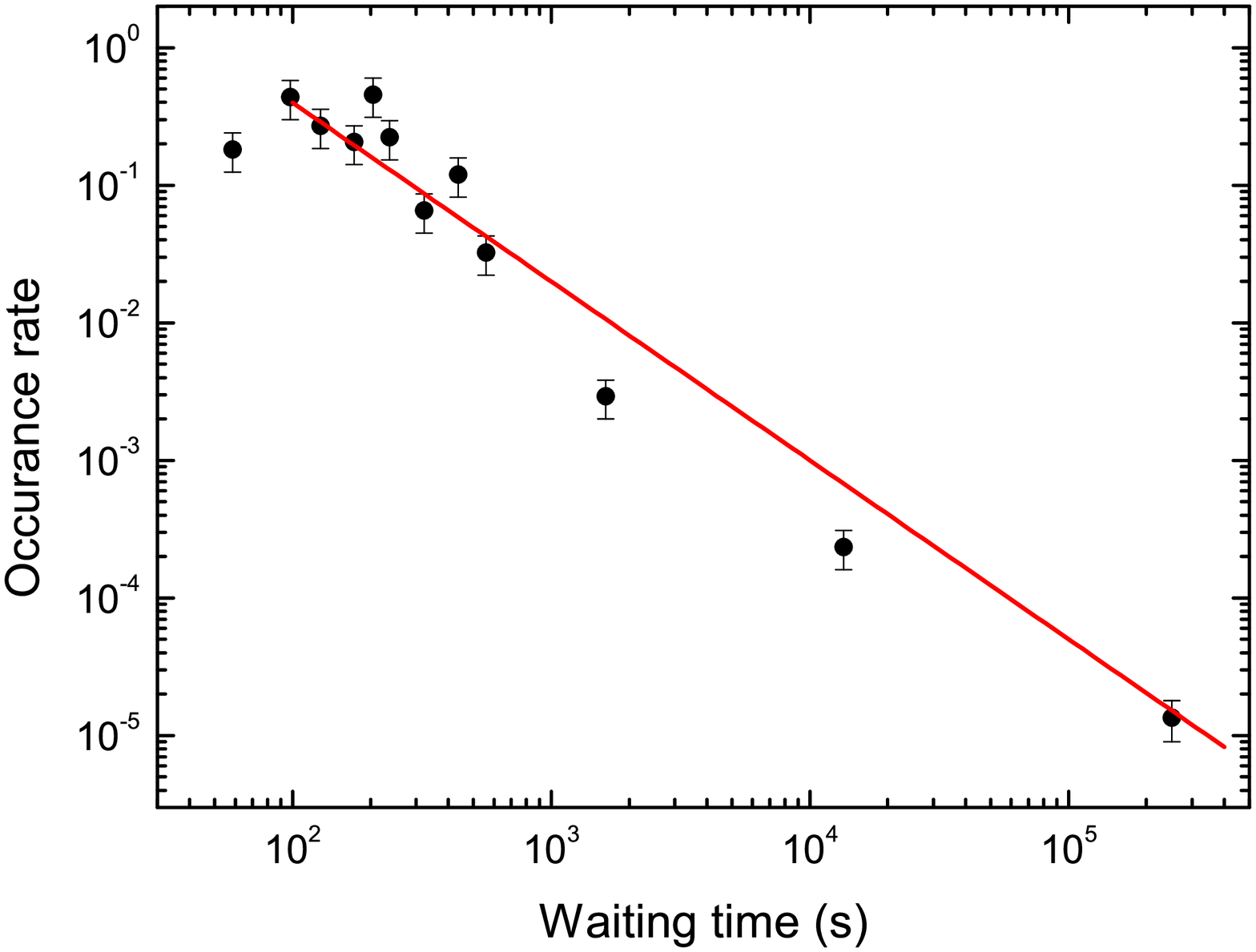}
\includegraphics[angle=0,scale=0.30]{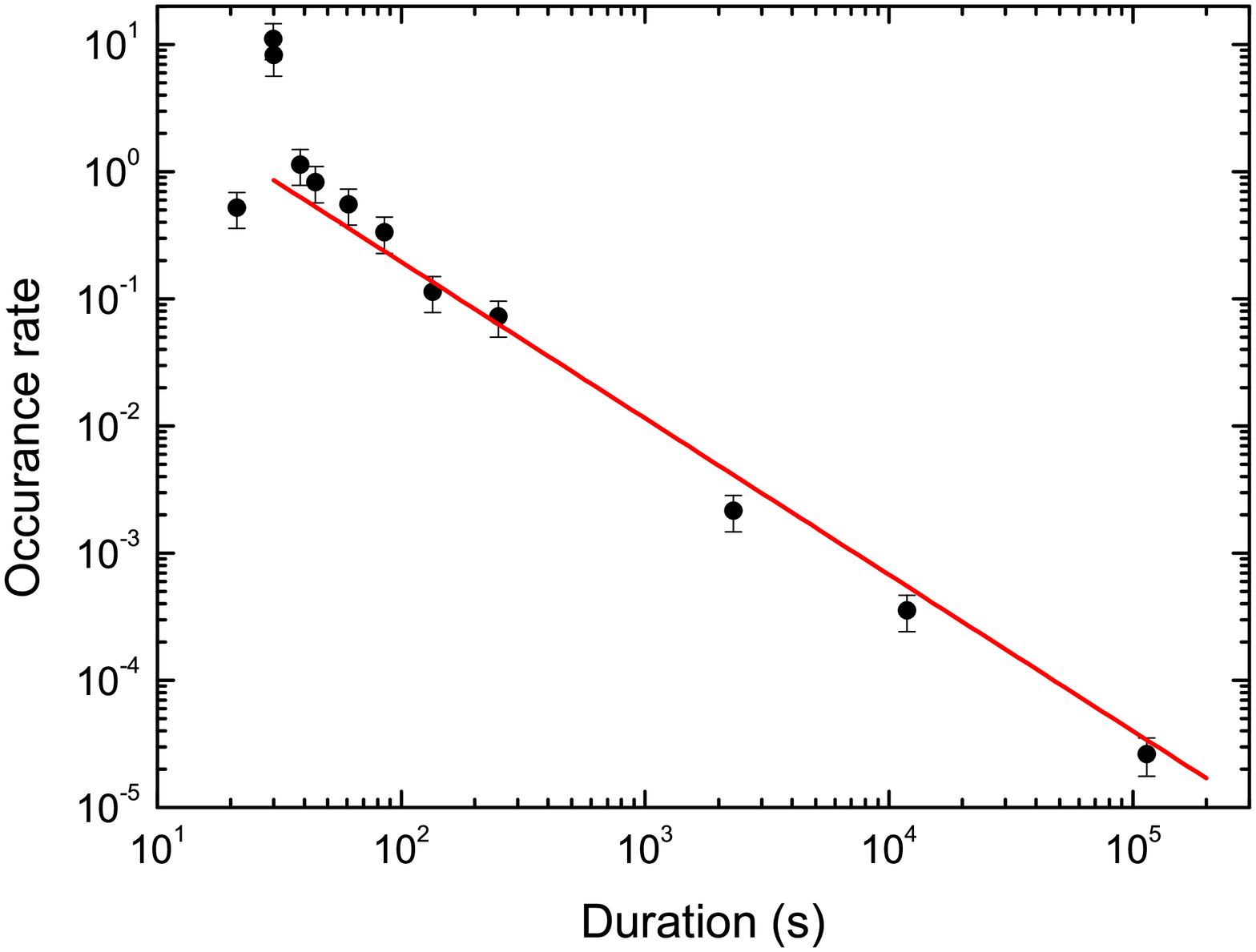}
\includegraphics[angle=0,scale=0.30]{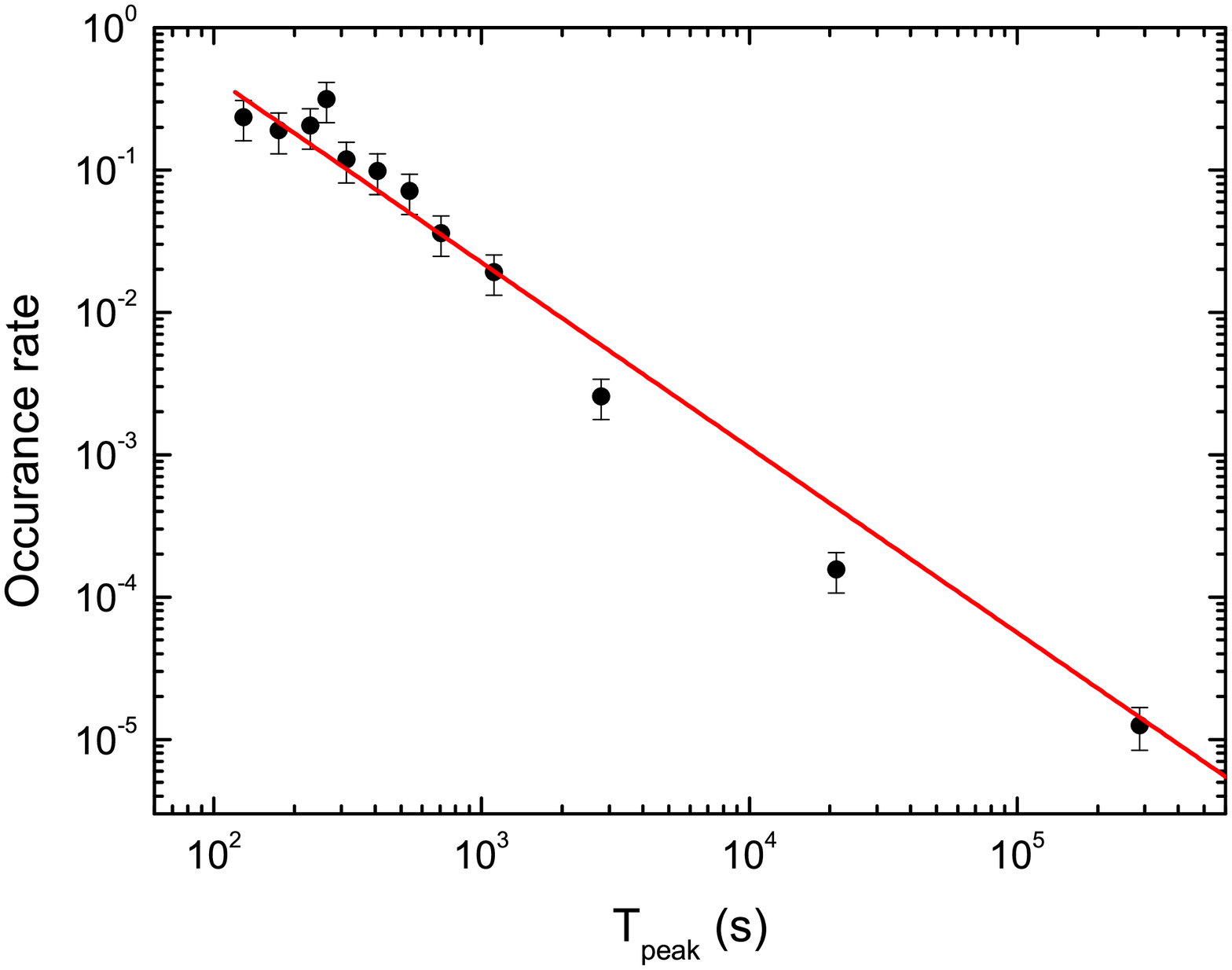}
\includegraphics[angle=0,scale=0.30]{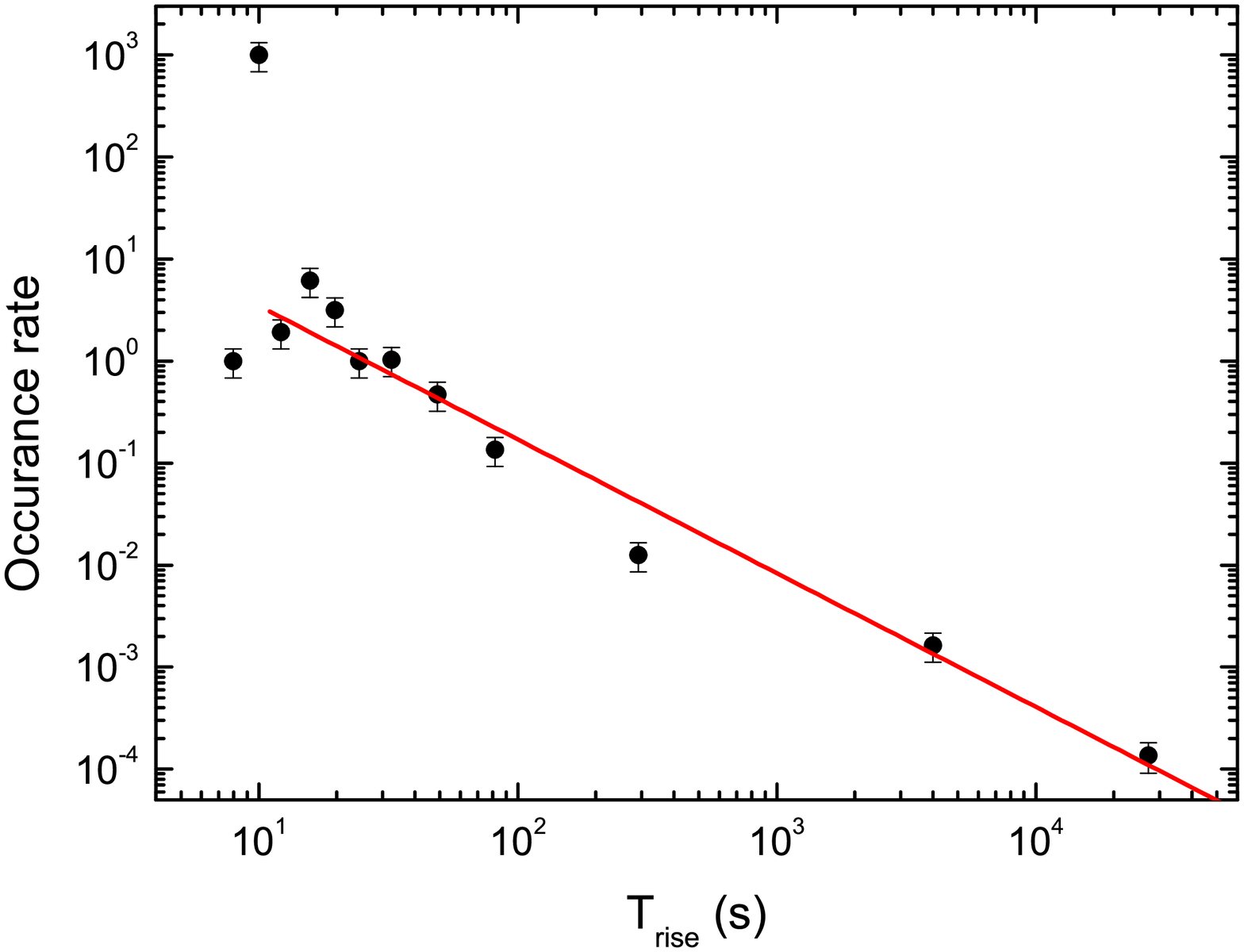}
\includegraphics[angle=0,scale=0.30]{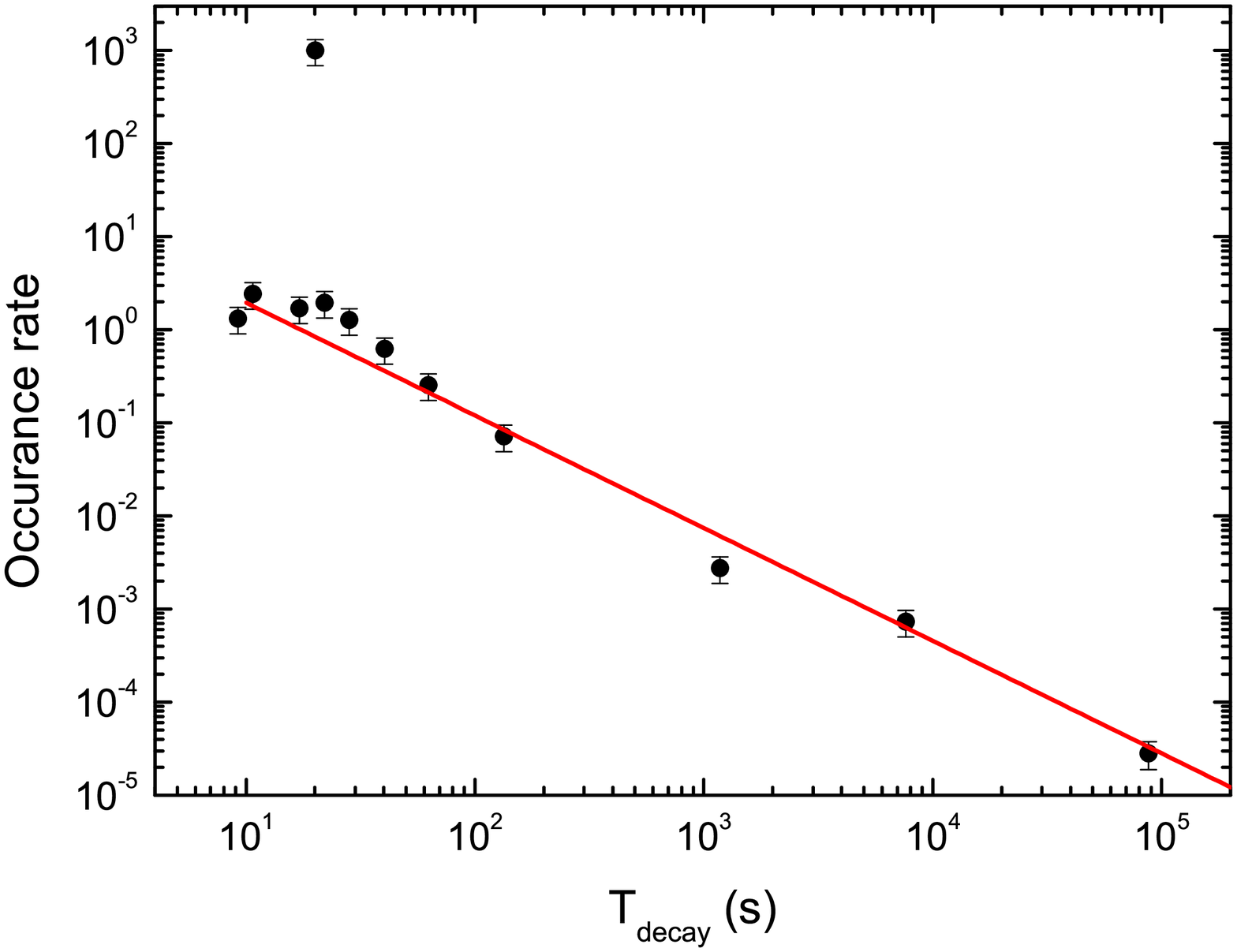}
\caption{The differential distributions of GRB optical flares. 119
GRB optical flares are used. The best-fitting indices for the
differential distributions of the waiting time, duration time, peak
time, rise time and decay time of the optical flares are
$1.24\pm0.08$, $1.23\pm0.07$, $1.28\pm0.09$, $1.31\pm0.10$ and
$1.21\pm0.07$, respectively. }
\end{figure*}

\begin{figure*}
\includegraphics[angle=0,scale=0.30]{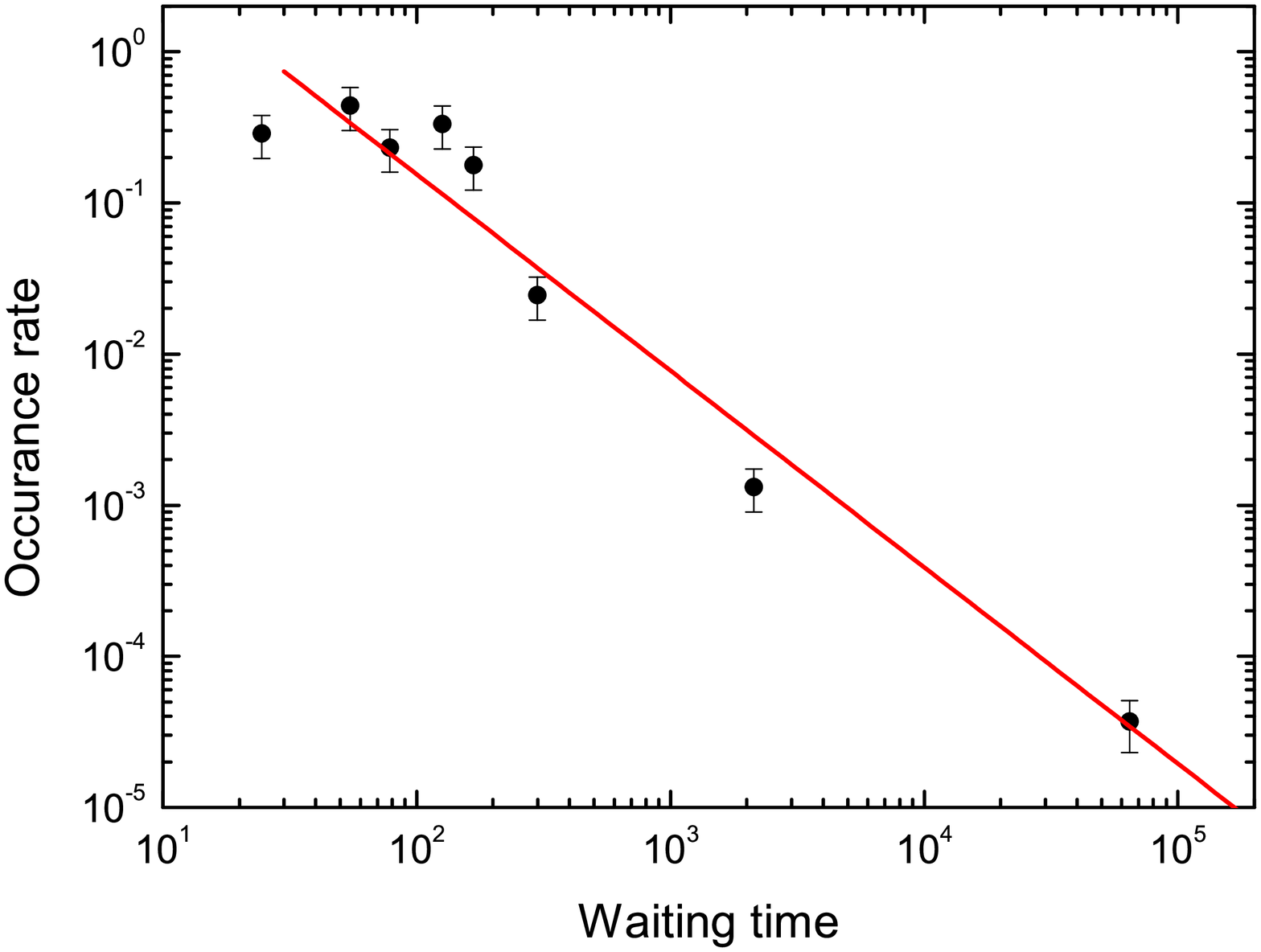}
\includegraphics[angle=0,scale=0.30]{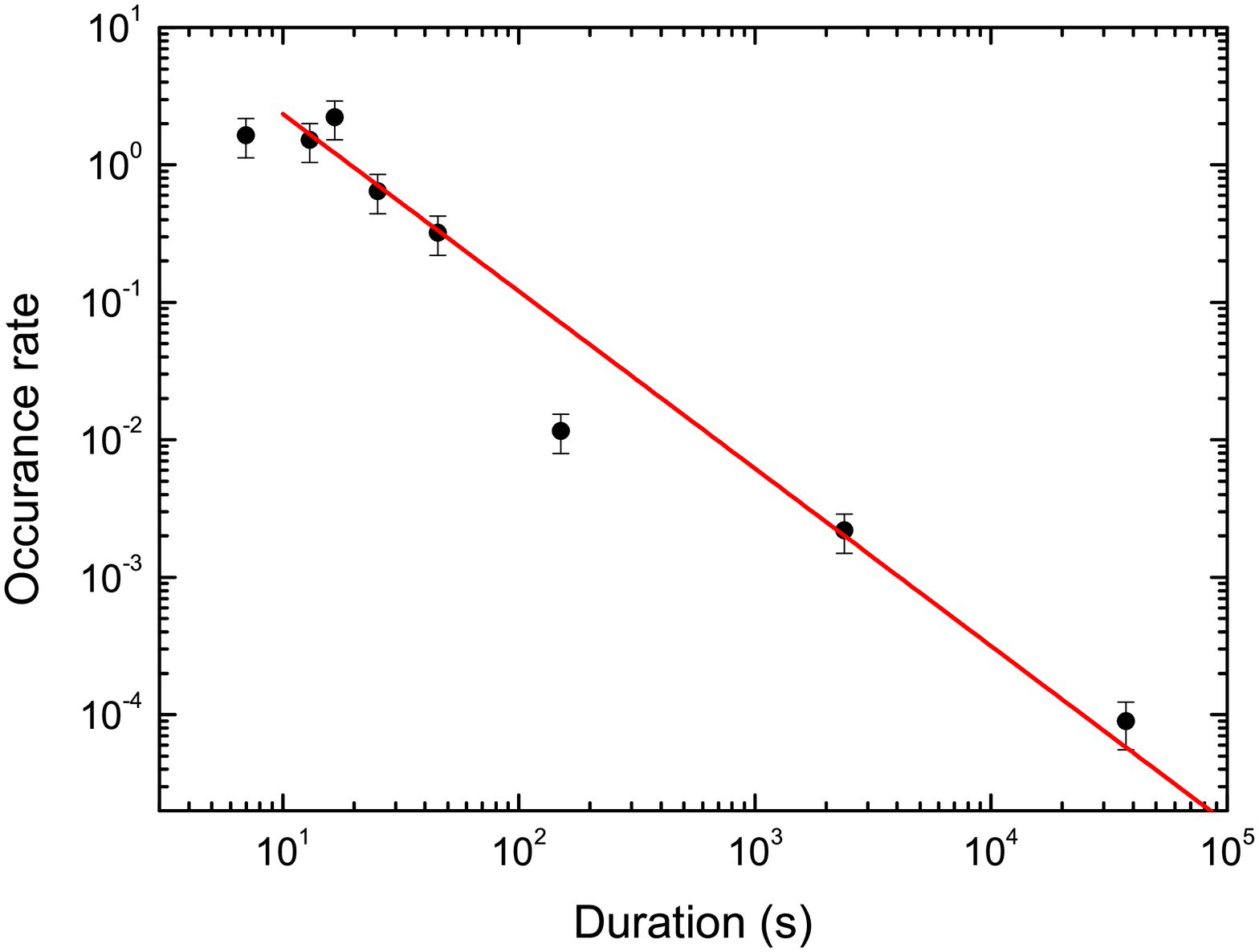}
\includegraphics[angle=0,scale=0.30]{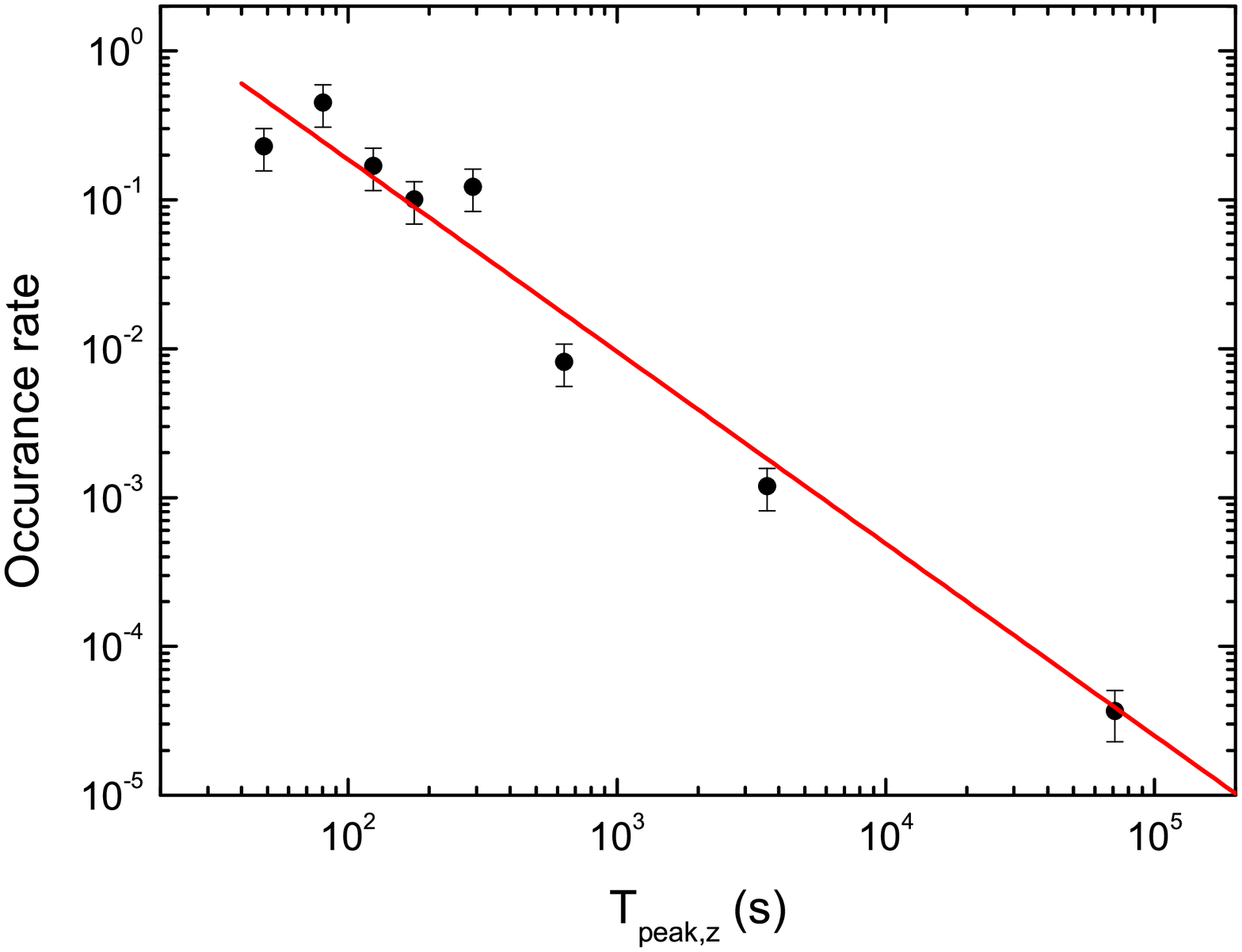}
\includegraphics[angle=0,scale=0.30]{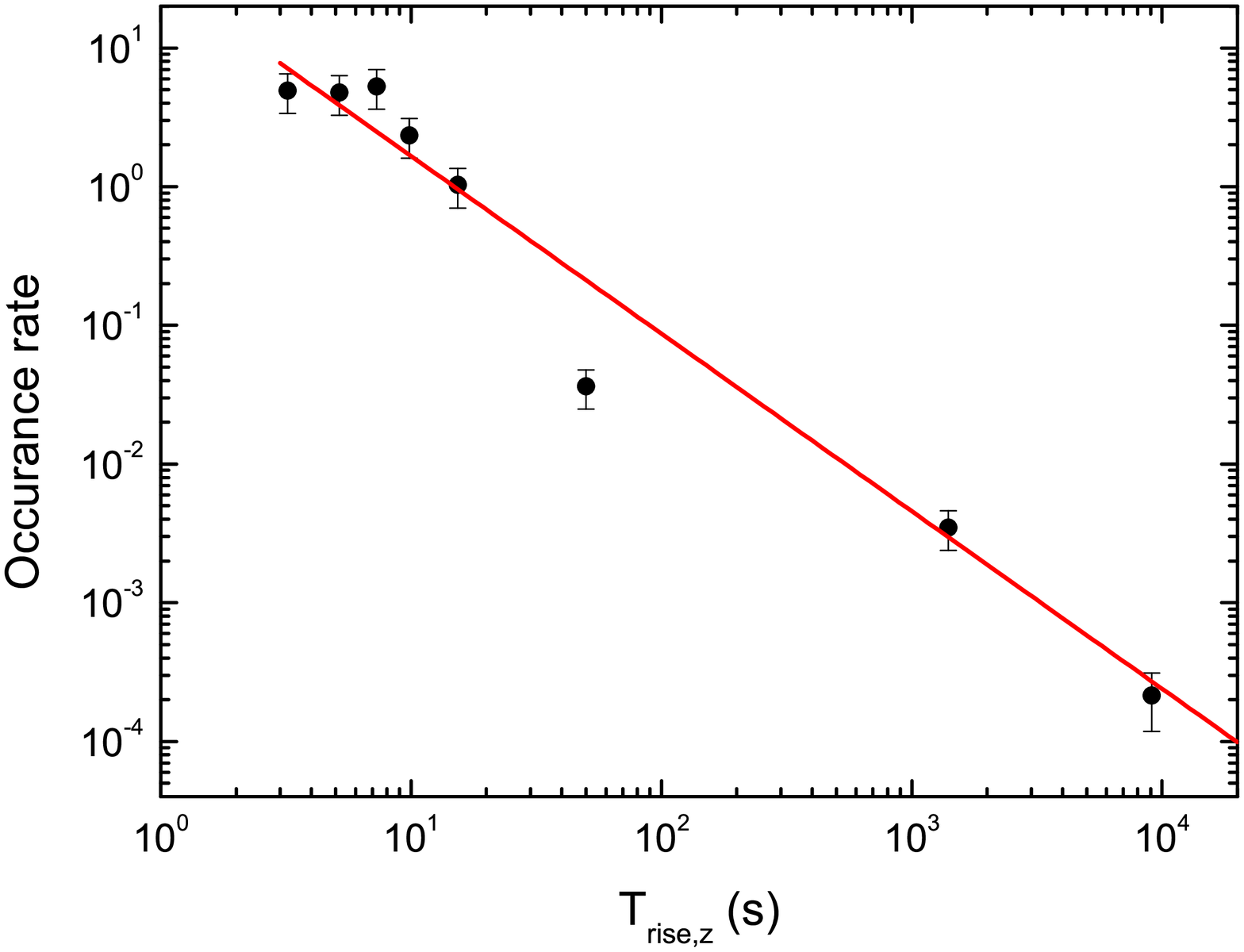}
\includegraphics[angle=0,scale=0.30]{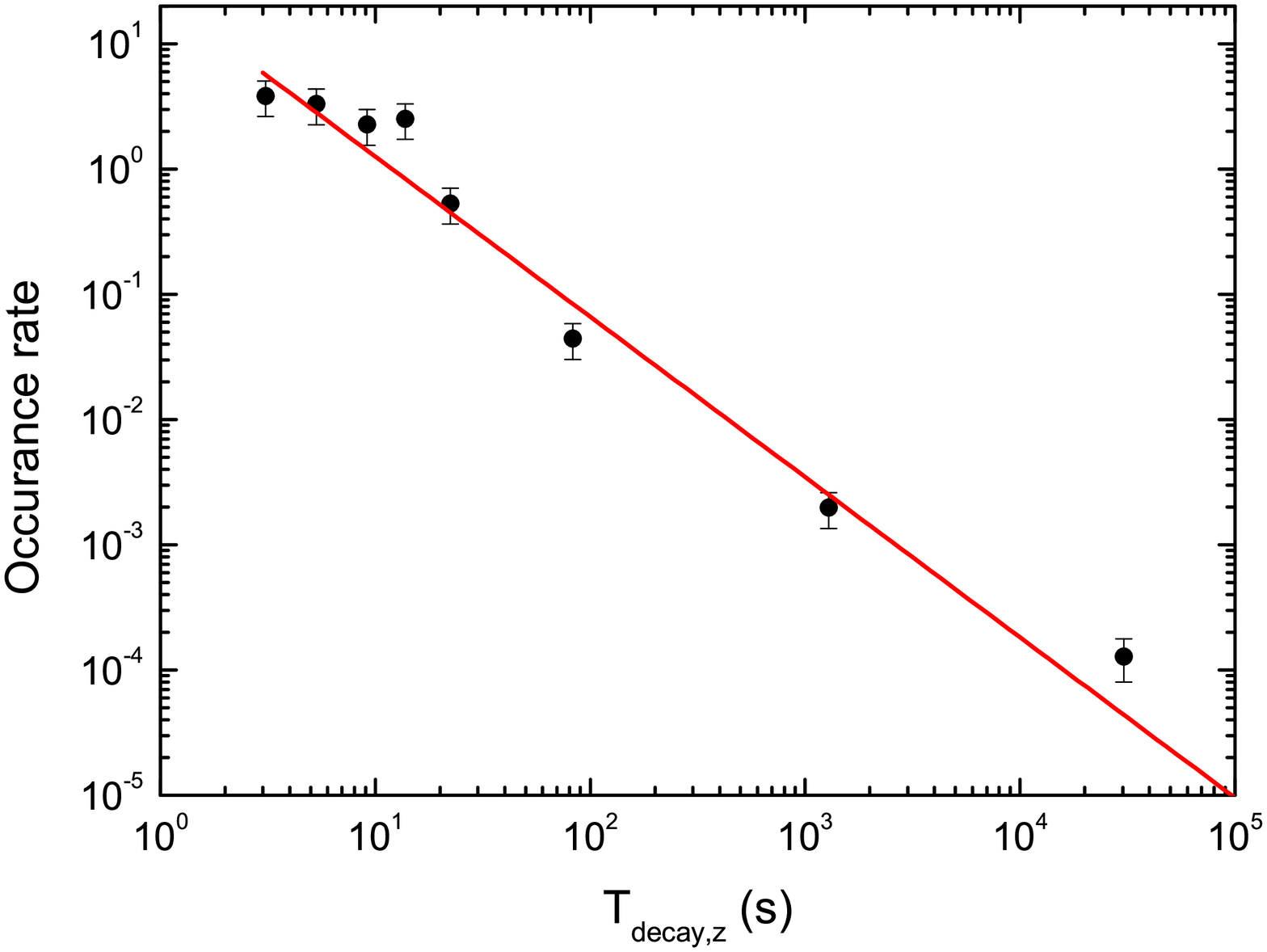}
\caption{The differential distributions of 77 GRB optical flares
with redshifts. The best-fitting indices for the differential
distributions of the waiting time, duration time, peak time, rise time
and decay time of optical flares are $1.30\pm0.11$, $1.29\pm0.09$,
$1.29\pm0.10$, $1.27\pm0.10$ and $1.28\pm0.11$, respectively.}
\end{figure*}


\clearpage
\begin{deluxetable}{ccccccccccccccccccccccccc}
\tabletypesize{\scriptsize} \tablecaption{Results of the linear
regression analysis for optical flares. $R$ is the Spearman
correlation coefficient, $P$ is the chance probability, and $\delta$
is the correlation dispersion.} \tablewidth{0pt}

\tablehead{ \colhead{Correlations}& \colhead{Expressions}&
\colhead{$R$}& \colhead{$P$}& \colhead{$\delta$} }

\startdata
\hline
$T_{decay}(T_{rise})$ & $\log T_{decay}=(0.17\pm 0.11)+(0.99\pm0.05)\times \log T_{rise}$ & 0.87 & $<10^{-4}$ & 0.57 \\
$T_{duration}(T_{peak})$ & $\log T_{duration}=(-1.00\pm 0.15)+(1.11\pm0.05)\times \log T_{peak}$ & 0.91 & $<10^{-4}$ & 0.45\\
$T_{peak}(T_{waiting})$ & $\log T_{peak}=(0.47\pm 0.07)+(0.91\pm0.03)\times \log T_{waiting}$ & 0.96 & $<10^{-4}$ & 0.26 \\
$T_{duration}(T_{waiting})$ & $\log T_{duration}=(-0.56\pm 0.14)+(1.05\pm0.05)\times \log T_{waiting}$ & 0.90 & $<10^{-4}$ & 0.48\\
\enddata
\end{deluxetable}



\end{document}